\documentclass[pre,twocolumn,showpacs,superscriptaddress]{revtex4}

\usepackage{graphicx}
\usepackage{amssymb}
\usepackage{amsmath}
\usepackage{amsfonts}
\usepackage{mathrsfs}
\usepackage[usenames]{color}
\definecolor{mygrey}{gray}{0.75}

\newcommand {\be} {\begin{equation}}   
\newcommand {\ee} {\end{equation}}
\newcommand {\bea} {\begin{eqnarray}}
\newcommand {\eea} {\end{eqnarray}}
\newcommand {\bes} {\begin{displaymath}}
\newcommand {\ees} {\end{displaymath}}
\newcommand {\beas} {\begin{eqnarray*}}
\newcommand {\eeas} {\end{eqnarray*}}

\begin{document}

\title{Master equation approach to DNA-breathing in heteropolymer DNA}

\author{Tobias Ambj{\"o}rnsson}
\email{ambjorn@nordita.dk}
\affiliation{NORDITA (Nordic Institute for
Theoretical Physics), Blegdamsvej 17, DK-2100 Copenhagen \O, Denmark.}
\affiliation{Present address: Department of Chemistry,
Massachusetts Institute of Technology, 77 Massachusetts Avenue, Cambridge,
MA 02139, USA}
\author{Suman K. Banik}
\affiliation{Dept. of Physics, Virginia Polytechnic Institute and State
University, Blacksburg, VA 24061-0435, USA}
\author{Michael A. Lomholt}
\affiliation{NORDITA (Nordic Institute for
Theoretical Physics), Blegdamsvej 17, DK-2100 Copenhagen \O, Denmark.}
\affiliation{Present address: Department of Physics, University
of Ottawa, 150 Louis Pasteur, Ottawa, Ontario  K1N 6N5, Canada}
\author{Ralf Metzler}
\email{metz@nordita.dk}
\affiliation{NORDITA (Nordic Institute for
Theoretical Physics), Blegdamsvej 17, DK-2100 Copenhagen \O, Denmark.}
\affiliation{Present address: Department of Physics, University
of Ottawa, 150 Louis Pasteur, Ottawa, Ontario  K1N 6N5, Canada}

\begin{abstract}
After crossing an initial barrier to break the first base-pair (bp) in
double-stranded DNA, the disruption of further bps is characterized by
free energies between less than one to a few $k_BT$. This causes the
opening of intermittent single-stranded bubbles. Their unzipping and
zipping dynamics can be monitored by single molecule fluorescence or
NMR methods. We here establish a dynamic description of this DNA-breathing
in a heteropolymer DNA in terms of a master equation that governs the
time evolution of the joint probability
distribution for the bubble size and position along the sequence. The
transfer coefficients are based on the Poland-Scheraga free energy model.
We derive the autocorrelation function for the bubble dynamics and the
associated relaxation time spectrum. In particular, we show how one can
obtain the probability densities of individual bubble lifetimes and of the
waiting times between successive bubble events from the master equation.
A comparison to results of a stochastic Gillespie simulation shows
excellent agreement.
\end{abstract}


\maketitle 

\section{Introduction}

Even at room temperature the DNA double-helix opens up locally to form
intermittent flexible single-stranded domains, the \emph{DNA-bubbles}.
Their size increases from a few broken base-pairs (bps) to a few hundred
open bps just below the melting temperature $T_m$, until a transition
occurs to full denaturation \cite{kornberg,poland,wartell,cantor_schimmel}.
DNA stability is effected by combination of the free energies $\epsilon_{\rm
hb}$ for breaking a Watson-Crick hydrogen bond between complementary AT and
GC bases in a single bp, and the ten independent stacking free energies
$\epsilon_{\rm st}$ for disrupting the interactions between a nearest neighbor
pair of bps. At 100 mM NaCl concentration and $T=37^\circ$C the hydrogen
bonding amounts to $\epsilon_{\rm hb}=1.0 k_BT$ for a single AT and $0.2 k_BT$
for a GC-bond \cite{REM}. The weakest (strongest) stacking energies was found
to be the TA/AT (GC/CG) with free energies $\epsilon_{\rm st}=-0.9 k_BT$
($-4.1 k_BT$). Note that negative values for the free energies denote 
stable states. The relatively small free energies for base stacking stem
from the fact that relatively large amounts of binding enthalpy on the one
hand, and entropy release on breaking the stacking interactions and
Watson-Crick bonds on the other hand, almost
cancel. Bubble initiation, in contrast, is characterized by breaking of two
stacking interactions with the first bp, whose enthalpy cost cannot be
balanced by the two, still strongly confined, liberated bases. Thus, the
creation of two boundaries between the intact double-helix and the bubble
nucleus is associated with an activation cost of some 7 to 12 $k_BT$
corresponding to the Boltzmann factor $\sigma_0\simeq 10^{-3}\ldots
10^{-5}$ \cite{poland,wartell,blake,blossey}. The cooperativity parameter
$\sigma_0$ is related to
the ring-factor $\xi$ used in \cite{FK,FK1}, see below. The high bubble
initiation barrier indeed guarantees the stability of DNA at physiological
conditions. Bubble opening occurs predominantly in zones rich in the
weaker AT bps; with increasing $T$, they spread to regions with progressively
higher GC content \cite{wartell,poland,kornberg}. Thermally driven,
a DNA-bubble fluctuates in size by relative random motion of the two zipper
forks. In addition to observations using NMR techniques \cite{gueron},
this \emph{DNA-breathing} was recently monitored on the single
bubble level by fluorescence correlation methods, demonstrating a multistate
kinetics that reflects the stepwise unzipping and zipping of bps. The
lifetime of a bubble was shown to range up to a few ms \cite{altan}.

From a biology or biochemistry point of view, DNA-breathing is of interest,
as it is implicated to influence the binding of binding proteins, enzymes, and
other chemicals to DNA. Thus, the relatively fast bubble dynamics with respect
to the binding rates of proteins, that selectively bind to single-stranded
DNA, provides a kinetic block preventing DNA denaturation through these
proteins \cite{rich1,pant,pant1}. This was investigated in
detail on the bases of a homopolymer approach in Refs.~\cite{tobias,JPC}.
Similarly, the increase of the bubble formation probability as well as
of the bubble lifetime at certain places along the sequence is believed
to facilitate the initiation of transcription by RNA polymerase.
This latter point is studied in depth in Ref.~\cite{bj}.

In this paper we investigate a $(2+1)$-variable master equation, that
governs the time evolution of the probability distribution $P(m,x_L,t)$
to find a bubble consisting of $m$ broken bps with left fork position $x_L$
along the sequence. With this approach, that is, an arbitrary DNA
sequence can be analyzed, and its breathing behavior predicted. We
discuss the exact form of the transfer matrix containing the rate
coefficients for all permitted moves, and derive the bubble autocorrelation
function with associated relaxation time spectrum. To be able to connect
to the time series obtained from the complementary stochastic simulation,
we derive the probability densities according to which individual bubble
lifetimes and the time intervals between successive bubble events are
distributed. Finally, we show that in the homopolymer limit, analytical
results can be obtained.

\section{One bubble partition function and transfer coefficients}
\label{sec:General model}

\begin{figure}
\begin{center}
\includegraphics[width=6.8cm]{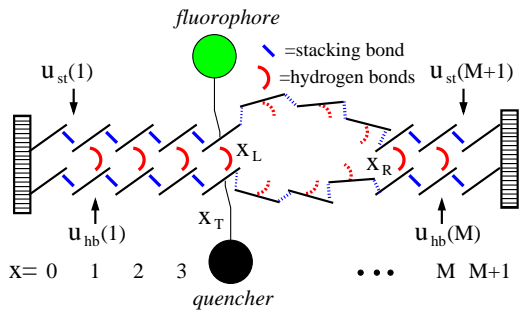}
\end{center}
\caption{Clamped DNA domain with internal bps $x=1$ to $M$,
statistical weights $u_{\rm hb}(x)$, $u_{\rm st}(x)$, and tag position
$x_T$. The DNA sequence enters through the statistical weights $u_{\rm
st}(x)$ and $u_{\rm hb}(x)$ for disrupting stacking and hydrogen bonds,
respectively. The bubble breathing process consists of the initiation
of a bubble and the subsequent motion of the forks at positions $x_L$
and $x_R$, see also Fig. \ref{fig:rates}.}
\label{fig:bubbles}
\end{figure}

Below the melting temperature $T_m$, a single bubble can be considered
to be statistically independent due to the high nucleation barrier for
initiating a bubble quantified by $\sigma_0\ll 1$ \cite{JPC}, such that
opening and merging of multiple bubbles are rare, and a one-bubble
picture is appropriate. In the particular case of the bubble constructs
used in the fluorescence correlation experiments of Ref.~\cite{altan},
the sequence is designed such that there is a single bubble domain.
Referring to these constructs, we consider a segment of double-stranded
DNA with $M$ internal bps. These bps are clamped at both ends such that
the bps $x=0$ and $x=M+1$ are always closed (Fig.~\ref{fig:bubbles}).
The sequence of bps determines the Boltzmann weights $u_{\rm hb}(x)=\exp
\{\epsilon_{\rm hb}(x)/(k_BT)\}$ for Watson-Crick hydrogen bonding at
position $x$, and the Boltzmann factor $u_{\rm st}(x)=\exp\{\epsilon_{\rm st}
(x)/ (k_BT)\}$ for pure bp-bp stacking between bps $x-1$ and $x$, respectively.
In the bubble domain, the left and right zipper fork positions $x_L$ and
$x_R$ denoting the right- and leftmost closed bp of the bubble are
stochastic quantities, whose random motion underlies the bubble dynamics.

Instead of using the fork positions $x_L$ and $x_R$, we prefer to work
with the left fork position $x_L$ and the bubble size $m=x_R-x_L-1$. For
these variables, the partition function of the bubble becomes
\be
\mathscr{Z}(x_L,m)=\frac{\xi'}{(1+m)^c}
\prod_{x=x_L+1}^{x_L+m} \hspace*{-0.2cm}u_{\rm hb}(x)
\prod_{x=x_L+1}^{x_L+m+1}\hspace*{-0.2cm}u_{\rm st}(x)
\label{part}
\ee
for $m\ge 1$. At $m=0$, we define $\mathscr{Z}(m=0)=1$. In relation
(\ref{part}), instead of the usual cooperativity parameter $\sigma_0$
we use the factor $\xi'=2^c\xi$ related to the ring factor $\xi\approx
10^{-3}$ introduced in Ref.~\cite{FK}. For a homopolymer, this $\xi$ is
related to $\sigma_0$ by $\sigma_0=\xi \exp\{\epsilon_{\mathrm{st}}/(k_BT)\}$
\cite{FK}. The denominator in Eq.~(\ref{part}) represents the entropy
loss on formation of a closed polymer loop, where the offset by one
accounts for finite size effects \cite{blake,fixman}. The associated
critical exponent is $c=1.76$ \cite{richard}. For a given bubble size,
the partition (\ref{part}) counts $m$ contributions from broken hydrogen
bonds and $m+1$ from disrupted stacking interactions. The partition
(\ref{part}) defines the equilibrium probability
  \be
P^{\rm eq}(x_L,m)=\frac{\mathscr{Z}(x_L,m)}{\mathscr{Z}(0)+\sum_{m=1}^M
\sum_{x_L=0}^{M-m}\mathscr{Z}(x_L,m)}
\label{eq:P_eq}
  \ee
for finding a bubble of size $m$ at location $x_L$.

The two variables $m$ and $x_L$ are the slow variables of the breathing
dynamics, while the polymeric degrees of freedom of the relatively
small bubble enter effectively through the partition (\ref{part}).
We moreover assume that the bubble is always close to thermal equilibrium
such that the partition (\ref{part}) defines the transition probabilities
between different states. These conditions allow us to introduce the
master equation (\ref{eq:master_eq_main}) with its transfer matrix
$\mathscr{W}$ below. To introduce the underlying time scales, we first
define the transfer coefficients.

\begin{figure}
\begin{center}
\includegraphics[width=8.8cm]{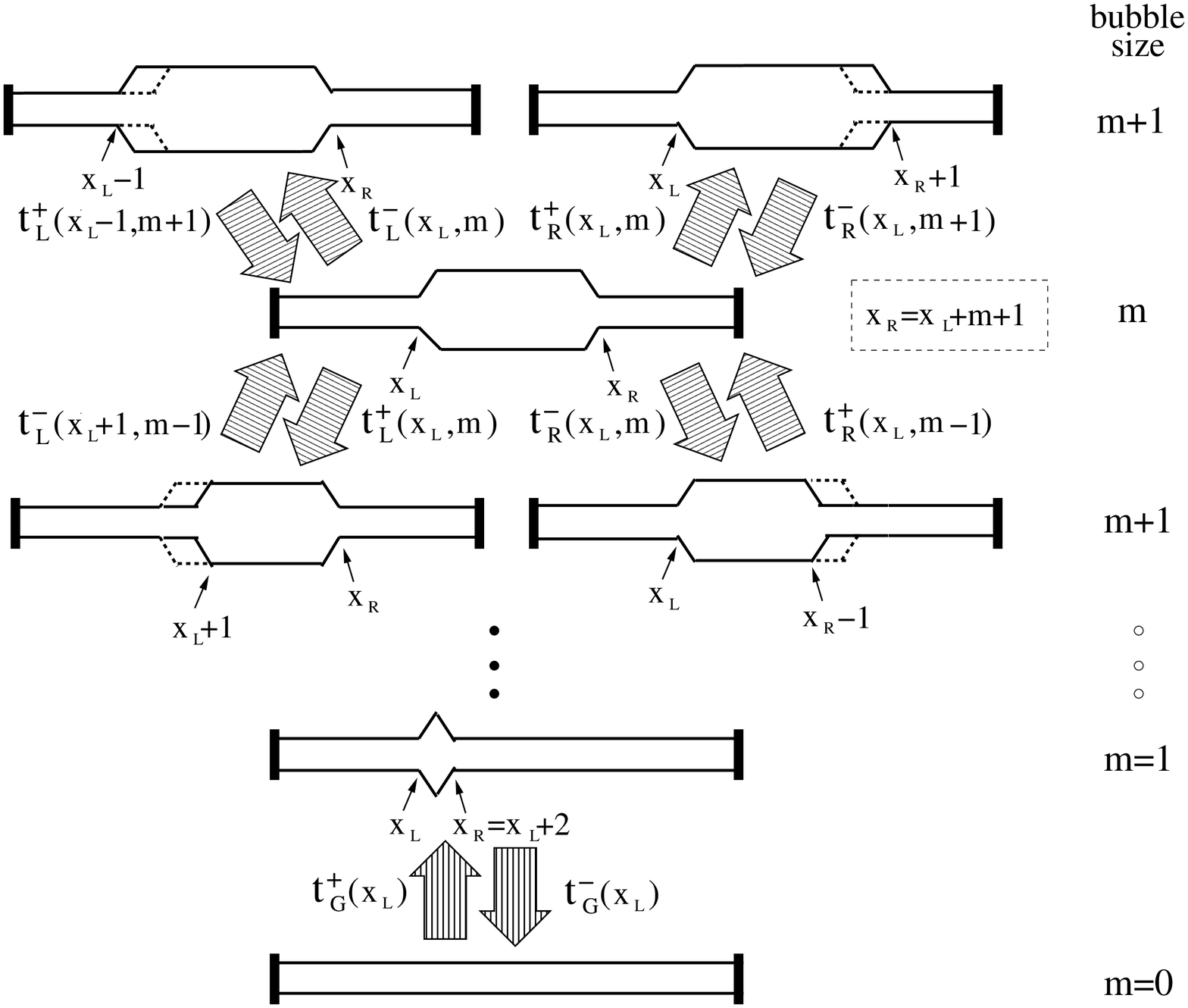}
\end{center}
\caption{Possible bubble (un)zipping transitions:
for $m\ge 2$, the four transfer rates $\mathsf{t}^{\pm}_{L/R}(x_L,m)$
completely determine the transitions \emph{out of} this state; the coefficients
$\mathsf{t}_L^{\pm}(x_L\mp 1,m\pm 1)$ and $\mathsf{t}_R^{\pm}(x_L,m\mp 1)$
specify the possible jumps \emph{into} this state. Jumps between state $m=1$
and ground state $m=0$ are described by $\mathsf{t}^+_G(x_L)$ and $\mathsf{t}^-
_G(x_L)$.}
\label{fig:rates}
\end{figure}

\begin{figure}
\begin{center}
\includegraphics[width=8.8cm]{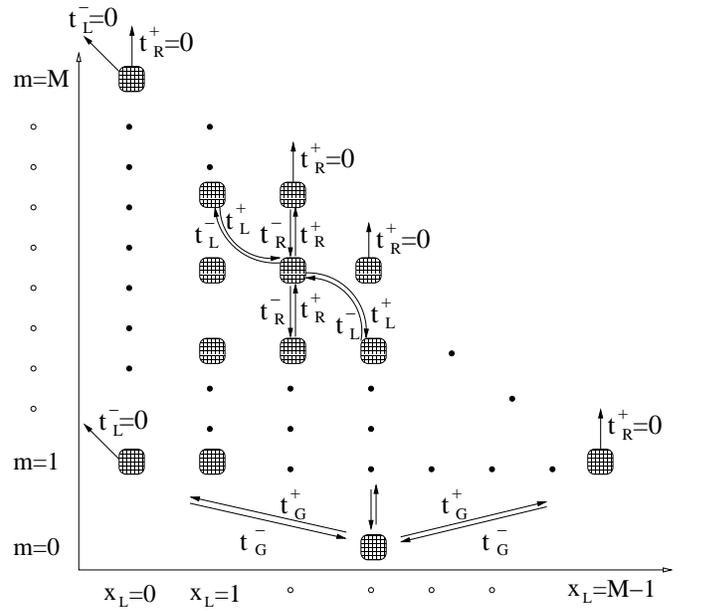}
\end{center}
\caption{Schematic representation of the $(x_L,m)$-lattice on which the
  DNA-bubble jump process takes place, with the permitted transitions;
  compare to Figs.~\ref{fig:bubbles} and \ref{fig:rates}. The boundary
  conditions Eq.~(\ref{eq:reflecting1}) are also indicated.}
\label{fig:lattice}
\end{figure}

The allowed transitions with the associated transfer (rate) coefficients
are sketched in Fig. \ref{fig:rates}. The left zipper fork is characterized
by the rate $\mathsf{t}^+_L(x_L,m)$ corresponding to the process $x_L
\rightarrow x_L+1$ of bubble size decrease, and $\mathsf{t}^-_L(x_L,m)$ for
$x_L\rightarrow x_L-1$ (bubble size increase). Similarly, we introduce
$\mathsf{t}^+_R(x_L,m)$ for $x_R \rightarrow x_R+1$ (bubble size increase)
and $\mathsf{t}^-_R(x_L,m)$ for $x_R \rightarrow x_R-1$ (decrease). These
rates are valid for transitions between states with $m\ge 1$. Bubble opening
(initiation) $m=0\rightarrow m=1$ is quantified by $\mathsf{t}_G^+
(x_L)$, and bubble closing (annihilation) $m=1\to0$ by $\mathsf{t}_G^-(x_L)$.
Note that by our definitions $\mathsf{t}_G^+(x_L)$ and $\mathsf{t}_G^-(x_L)$
actually correspond to bubble opening/closing at $x=x_L+1$. Clamping requires
that $x_L\ge 0$ and $x_R\le M+1$, corresponding to reflecting boundary
conditions
\cite{REM0}
  \be
\mathsf{t}_L^-(x_L=0,m)=\mathsf{t}_R^+(x_L,m=M-x_L)=0.
\label{eq:reflecting1}
  \ee  
In Fig.~\ref{fig:lattice} we sketch schematically the allowed transitions
in the $m$-$x_L$ plane.

In order to define the various transfer rates $\mathsf{t}$, we firstly
impose the detailed balance conditions (compare \cite{Risken,van_Kampen})
\begin{subequations}
\label{eq:det_balance}
\bea
\frac{\mathsf{t}_L^+(x_L-1,m+1)}{\mathsf{t}_L^-(x_L,m)}&=&
\frac{P_{\rm eq}(x_L,m)}{P_{\rm eq}(x_L-1,m+1)}\\
\frac{\mathsf{t}_R^-(x_L,m+1)}{\mathsf{t}_R^+(x_L,m)}&=&
\frac{P_{\rm eq}(x_L,m)}{P_{\rm eq}(x_L,m+1)}\\
\frac{\mathsf{t}_G^+(x_L)}{\mathsf{t}_G^-(x_L)}&=&
\frac{P_{\rm eq}(x_L,1)}{P_{\rm eq}(0)},
\eea
\end{subequations}
that ensure relaxation toward the equilibrium distribution $P^{\rm eq}(x_L,m)$,
see Eq.~(\ref{eq:P_eq}). The latter can be seen by recalling that the
equilibrium distribution (\ref{eq:P_eq}) is based on the Boltzmann
factors $u_{\mathrm{hb}}$ and $u_{\mathrm{st}}$ through the partition
(\ref{part}). The detailed balance
conditions do not uniquely define the transfer rates $\mathsf{t}$,
leaving a certain freedom of choice \cite{van_Kampen}. We use the
following conventions.

To define the zipping rate, we assume that it is independent of the
position $x$ along the DNA sequence. The picture behind is that the
closure of the bp is dominated by the requirement that the two bases
diffuse in real space until mutual encounter and eventual bond formation.
As sterically AT and GC bps are very similar, the zipping rate should not
significantly vary with the individual nature of the involved bps, and we
choose the constant rate $k/2$, see below. This rate $k$ is the only
adjustable parameter of our model, and has to be independently determined
from experiment or more fundamental models. This choice, as mentioned
above, is not unique. Instead, an $x$-dependence of $k$ could easily
be introduced by choosing different powers of the statistical weights
entering the rate coefficients such that they still fulfill detailed balance.

Decrease in bubble size due to zipping close of the bp closest to either the
left or the right zipper fork is therefore ruled by
  \be
\mathsf{t}^+_L(x_L,m)|_{m\ge 2}=t^-_R(x_L,m)|_{m\ge 2}=\frac{k}{2},
\label{eq:t_L_plus}
  \ee
respectively. The factor $1/2$ is introduced for consistency with previous
approaches \cite{tobias,JPC}. Note that for simplicity we do not introduce
the hook exponent discussed in previous studies \cite{PRL,JPC,bj}. This exponent
should be important for large bubbles, when during the zipping process not
only the bp at the zipper fork is moved, but also part of the vicinal
single-strand is dragged or pushed
along \cite{Di_Marzio_Guttman_Hoffman,PRL,JPC}.

Increase in bubble size is controlled by
\begin{subequations}
\begin{eqnarray}
\label{eq:t_L_minus_et}
\mathsf{t}_{L}^{-}(x_L,m)&=&\frac{k}{2}u_{\rm st}(x_L) u_{\rm hb}(x_{L})s(m),\\
\mathsf{t}_{R}^{+}(x_L,m)&=&\frac{k}{2}u_{\rm st}(x_R+1) u_{\rm hb}(x_{R})s(m),
\label{eq:t_L_minus_etc}
\end{eqnarray}
\end{subequations}
for $m\ge 1$, where
\be
s(m)=\left(\frac{1+m}{2+m}\right)^c.
\ee
For $m\ge 1$ we thus take the bubble increase rate coefficients proportional
to the first power of the Arrhenius factor $u_{\rm st}u_{\rm hb}=\exp\{
(\epsilon_{\mathrm{hb}}+\epsilon_{\mathrm{st}})/[k_BT] \}$ times the loop
correction $s(m)$. We stress that Eqs.~(\ref{eq:t_L_minus_et}) and
(\ref{eq:t_L_minus_etc}) are dictated by the detailed balance condition,
once the convention (\ref{eq:t_L_plus}) is established. As noted, detailed
balance would still be fulfilled, for instance, if
only a fractional power $\alpha^q$ of the Arrhenius factor $\alpha$ appeared
in the opening rates if complemented by the respective power $\alpha^{1-q}$
in the closing rates.

Finally, we define
\begin{subequations}
\bea
\label{init}
\mathsf{t}_G^+(x_L)&=& k \xi's(0)u_{\rm st}(x_L+1)u_{\rm hb}(x_L+1)\\
\nonumber
&& \times u_{\rm st}(x_L+2)\\
\mathsf{t}_G^-(x_L)&=&k
\label{eq:t_G_plus}
\eea
\end{subequations}
for bubble initiation and annihilation from and to the zero-bubble state $m=0$,
with the bubble initiation factor $\xi'$ in the expression for $\mathsf{t}_G^
+$. As bubble initiation involves breaking of two stacking interactions at
consecutive bps, we have the factor $u_{\rm st}(x_L+1)u_{\rm st}(x_L+2)$ in
expression (\ref{init}). The last open bp can zip close from either side, so
the bubble closing rate $\mathsf{t}_G^-(x_L)$ makes up twice the zipping rate
of a single fork.

The rates $\mathsf{t}$ together with the boundary conditions fully
determine the bubble dynamics. In the next Section, we establish the
master equation for the time evolution of the distribution $P(m,x_L,t)$
and derive the associated quantities.

\section{Master equation for DNA-breathing}
\label{sec:ME}

The joint probability distribution $P(t)=P(x_L,m,t;x_L',m',0)$ measures the
likelihood that at time $t$ the system is in state $\{x_L,m\}$ and that at
$t=0$ it was in state $\{x_L',m'\}$. Its time evolution is controlled by the
master equation
  \be
\frac{\partial}{\partial t}P(t)= \mathscr{W} P(t).
\label{eq:master_eq_main}
  \ee
The explicit form of the transfer matrix $\mathscr{W}$ is discussed in
detail in Sec.~\ref{sec:ME_details}. Here, we concentrate on how to derive
the quantities relevant for the description of DNA-breathing experiments.
In that course we introduce the eigenmode ansatz \cite{van_Kampen,Risken}
  \be
P(t)=\sum_p c_p Q_p \exp(-\eta_p t),
\label{eq:spectral_decomp}
  \ee 
where the coefficients $c_p$ are fixed by the initial condition. Combining
Eqs.~(\ref{eq:spectral_decomp}) and (\ref{eq:master_eq_main}), the
eigenvalue equation
  \be
 \mathscr{W} Q_p=-\eta_p Q_p
 \label{eq:eigenvalue_eq}
  \ee
yields, compare Ref.~\cite{van_Kampen} and below for more details. The
eigenvalues $\eta_p$ and eigenvectors $Q_p$ allow one to compute
any quantity of interest. In fact, the autocorrelation function for bubble
breathing and the corresponding relaxation time distribution are quite
straightforward to obtain, see section \ref{seca}. Below, in section
\ref{secb}, we discuss the more subtle point how the
probability densities for the bubble lifetime and the interbubble event
waiting time can be derived.

\subsection{Blinking autocorrelation function of a tagged bp}
\label{seca}

Motivated by the fluorescence correlation setup in Ref.~\cite{altan} we
are interested in the state of a tagged bp at $x=x_T$, see
Fig.~\ref{fig:bubbles}. In the experiment fluorescence occurs if the bps in a
$\Delta$-neighborhood of the fluorophore position $x_T$ are open
\cite{altan}. Measured fluorescence blinking time series thus correspond to the
stochastic variable $I(t)$, defined by
\begin{equation}
I(t) = \left \{
\begin{array}{cc}
1 & \text{at least all bps in $[x_T-\Delta,x_T+\Delta]$ open} \\
0 & \text{otherwise},\\
\end{array}
\right..
\end{equation}
The stochastic variable is $I=1$ if the system is in the phase space region
defined by 
  \be
\mathbb{R}1:\{ 0 \leqslant x_L \leqslant x_T-\Delta-1, \ x_T-x_L+\Delta
\leqslant m \leqslant M-x_L \}.
  \ee
Conversely, $I=0$ corresponds to the complement $\mathbb{R}0$.

The equilibrium autocorrelation function of fluorescence blinking is
defined by
\be
A(x_T,t)=\langle I(t)I(0)\rangle -(\langle I \rangle) ^2,
\label{eq:A_t}
\ee
where the angles $\langle\cdot\rangle$ denote an ensemble average over
the equilibrium distribution $P_{\mathrm{eq}}$. $A(x_T,t)$ quantifies
the relaxation dynamics of the tagged bp. This is evident from the
identity
\be
\langle I(t)I(0)\rangle=\sum_{I=0}^1 \sum_{I'=0}^1 I \rho(I,t;I',0)I'=
\rho(1,t;1,0),
\label{eq:I_t}
\ee
where $\rho(1,t;1,0)$ is the survival probability that $I(t)=1$
and that $I(0)=1$. From the definition of the two regions $\mathbb{R}1$
and $\mathbb{R}0$ it follows that $\rho(1,t;1,0)$ yields from summation
of $ P(x_L,m,t;x_L',m',0)$ solely over region $\mathbb{R}1$:
\be
\rho(1,t;1,0)=\sum_{x_L,m,x_L',m'\in \mathbb{R}1} P(x_L,m,t;x_L',m',0).
\ee
Together with Eq.~(\ref{eq:I_t}), combined with the eigenmode decomposition
(\ref{eq:spectral_decomp}), and under the assumption that initially the system
is at equilibrium, we obtain
\be
P(x_L,m,0;x_L',m',0)=\delta_{mm'}\delta_{x_Lx_L'}P_{\rm eq}(x_L,m),
\ee
such that we can rewrite the autocorrelation function (\ref{eq:A_t}) in
the form
  \be
A(x_T,t)=\sum_{p\neq 0} \left[T_p(x_T)\right]^2
\exp(-t/\tau_p).
\label{eq:A_xT_main}
  \ee
Here, we use the relaxation times $\tau_p=1/\eta_p$, and abbreviate
\be
T_p(x_T)=\sum_{x_L=0}^{x_T-\Delta-1} \sum_{m=x_T-x_L+\Delta}^{M-x_L}
Q_p(x_L,m).
\label{eq:T_p}
\ee
In all illustrations, we plot the normalized form of the autocorrelation
function, $A(x_T,t)/A(x_T,0)=A(x_T,t)/\sum\left[T_p(x_T)\right]^2$.

The autocorrelation function $A(x_T,t)$ can be rewritten as the integral
$A(x_T,t)=\int d\tau\exp(-t/\tau) f(x_T,\tau)$ defining the weighted
spectral density (relaxation time spectrum)
density
\be
f(x_T,\tau)=\sum_{p\neq 0} [T_p(x_T)]^2 \delta(\tau-\tau_p).
\label{eq:f_tau}
\ee
This quantity indicates how many different exponential modes contribute 
to the autocorrelation function. If $f(x_T,\tau)$ is very narrow, the
process is approximately exponential, whereas a broad relaxation time
spectrum indicates that many different modes play together.

\subsection{Survival and waiting time densities of a tagged bp}
\label{secb}
\label{sec:ME_surv_wait}

The autocorrelation function $A(x_T,t)$ is an equilibrium average
measure for a single bubble. It does not contain any information
on how the lifetime of individual bubbles is distributed, nor
on the waiting time that elapsed between annihilation of a bubble
and the initiation of the next. This information is provided by
the survival time and waiting time densities $\phi(t)$ and $\psi(t)$.
Here we derive these two quantities.

Survival time and the waiting time densities correspond to a first passage
problem, to respectively start from an initial state with $I(0)=1$ or $I(0)
=0$, and transit to a state $I(t)=0$ or $I(t)=1$ after time $t$. To obtain
these quantities from the master equation framework, one needs to solve the
reduced eigenvalue problem \cite{van_Kampen}
 \be
\mathscr{W}\bar{Q}_p=-\bar{\eta}_p\bar{Q}_p
\label{eq:eigenvalue_eq_reduced2}
  \ee
for coordinates belonging to $\mathbb{R}1$ and $\mathbb{R}0$. Details are
collected in Sec.~\ref{sec:ME_details}. The reduced eigenvalue ans{\"a}tze
(\ref{eq:eigenvalue_eq_reduced2}) for $\mathbb{R}1$ and $\mathbb{R}0$
possess only positive eigenvalues, $ \bar{\eta}_p>0$ for all $p$. This
reflects the fact that there exist transitions from one region to the
other, such that probability "leaks out".
In terms of the reduced eigenvalues $\bar{\eta}_p$ and eigenvectors
$\bar{Q}_p$ the survival and waiting time densities become
\begin{subequations}
\label{eq:Psi-Phi}
  \bea
\psi(t)&=& \sum_{p\in \mathbb{R}0} \bar{\eta}_p \bar{c}_p \exp(-\bar{\eta}_p t)\\
\phi(t)&=& \sum_{p\in \mathbb{R}1} \bar{\eta}_p \bar{c}_p \exp(-\bar{\eta}_p t)
  \eea
with the coefficients
  \be
\bar{c}_p=\frac{\bar{\eta}_p [\sum_{x_L,m} \bar{Q}_p (x_L,m)]^2}{\sum_p
\bar{\eta}_p [\sum_{x_L,m} \bar{Q}_p (x_L,m)]^2}.
\label{eq:bar_c_p_main}
  \ee
\end{subequations}
The sums over $m,x_L$ are restricted to regions $\mathbb{R}1$ and $\mathbb{R}0$
for the survival and waiting time densities, respectively. Both survival and
waiting time probability densities are normalized, $\int\psi(t)dt=1$ and $\int
\phi(t)dt=1$, since $\sum_p\bar{c}_p=1$.

We point out that a non-trivial problem connected to obtaining the
appropriate expressions for $\psi(t)$ and $\phi(t)$ is how to choose
the right initial distribution of states (there are many states
corresponding to a bubble being just open/closed). We chose an initial
distribution determined by the distribution of stationary flux into
the regions $\mathbb{R}1$ and $\mathbb{R}0$. This choice guarantees that
(for long times) the ratio of the time spent in the $I=1$ state versus
the time spent in the
$I=0$ state is given by the equilibrium results as required by
ergodicity, see Sec.~\ref{sec:ME_details} for details. In appendix
\ref{sec:Gillespie} we briefly discuss how stochastic modeling can be used to
obtain single bubble time series, from which all quantities such as
fluorescence blinking autocorrelation function, as well as the survival and
waiting time densities can be distilled. Both approaches converge nicely
\cite{PRL,bj}.

\section{Master equation - the details}
\label{sec:ME_details}

In this Section we show the explicit form for the master equation with
its transfer matrix $\mathscr{W}$ and go into details of how to solve
it numerically. We also develop a formalism to derive the waiting and
survival densities $\psi(t)$ and $\phi(t)$.

\subsection{The $\mathscr{W}$-matrix}\label{sec:num_sol_master_eq}

\begin{figure}
  \begin{center}
    \includegraphics[width=6.8cm]{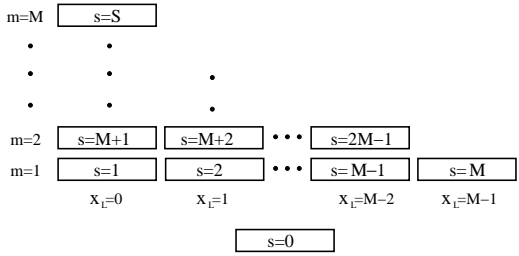}
  \end{center}
\caption{Enumeration scheme for the numerical analysis: The two-dimensional
grid points $(x_L,m)$ are replaced by a one-dimensional running variable $s$.
See text for details.}
\label{fig:s_space}
\end{figure}

In order to present an explicit expression for the
$\mathscr{W}$-matrix in Eqs.~(\ref{eq:master_eq_main}) and
(\ref{eq:eigenvalue_eq}) it is convenient to replace the two-dimensional
grid points $(x_L,m)$ by a one-dimensional coordinate $s$ counting all
lattice points, compare \cite{JPC}. We choose the enumeration
illustrated in figure \ref{fig:s_space}.
From this figure we notice that $m\in [0,M]$ and $x_L\in [0,M-m]$. We
label the ground state $m=0$ by $s=0$. For $m\ge 1$ an arbitrary
$s$-point can be obtained from a specific $(x_L,m)$ according to:
  \be
s=s|_{x_L}^m = (m-1)M-\frac{(m-1)(m-2)}{2}+x_L+1\label{eq:s}
  \ee
From this relation we notice that the maximum $s$ value is
  \be 
S={\rm max}\{ s\}=\frac{M(M+1)}{2},
  \ee
i.e., the size of the relevant $\mathscr{W}$-matrix (see below) scales
as $M^2$.  Expression (\ref{eq:s}) allows us to change the transfer
coefficients to the $s$-variable, $\mathsf{t}^\pm_{L/R}
(x_L,m)\rightarrow \mathsf{t}^\pm_{L/R} (s)$, using the explicit
expressions (\ref{eq:t_L_plus}), (\ref{eq:t_L_minus_et}), and
(\ref{eq:t_L_minus_etc}) for the transfer coefficients, together with the
boundary conditions in Eqs.~(\ref{eq:reflecting1}). Also $\mathsf{t}^{\pm}
_G(x_L,m)\to\mathsf{t}^{\pm}_G(s)$, following Eqs.~(\ref{init}) and
(\ref{eq:t_G_plus}). From Eq.~(\ref{eq:s})
and figure \ref{fig:s_space} we notice that
  \bea
s|_{x_L-1}^{m+1}&=&s|_{x_L}^m+M-m,\ {\rm for}\ x_L\ge 1 \ \& \ m\le M-1,
\nonumber\\
s|_{x_L+1}^{m-1}&=&s|_{x_L}^m-(M-m+1),\ {\rm for}\ m\ge 2,\nonumber\\
s|_{x_L}^{m-1}&=&s|_{x_L}^m-(M-m+2),\ {\rm for}\ m\ge 2,\nonumber\\
s|_{x_L}^{m+1}&=&s|_{x_L}^m+M-m+1,\nonumber\\
& & \ {\rm for}\ x_L\le M-(m+1) \ \& \ m\le M-1.\nonumber\\
  \eea
We can then write Eq.~(\ref{eq:eigenvalue_eq}) explicitly as
  \be
\sum_{s'}\mathscr{W}(s,s')Q_p(s')=-\eta_p Q_p(s),
\label{eq:eigen2}
  \ee
where the matrix-elements are
  \bea
\mathscr{W}(s,s+M-m)&=&\mathsf{t}_L^+(s+M-m),\nonumber\\
&& {\rm for} \ s  \pitchfork x_L\ge 1\,\& \, m>1\nonumber\\
\mathscr{W}(s,s-[M-m+1])&=&\mathsf{t}_L^-(s-[M-m+1]),\nonumber\\
&&  {\rm for} \ s \pitchfork m\ge 2 \nonumber\\
\mathscr{W}(s,s-[M-m+2])&=&\mathsf{t}_R^+(s-[M-m+2]),\nonumber\\
&&  {\rm for} \ s \pitchfork m\ge 2, \nonumber\\
\mathscr{W}(s,s+M-m+1)&=&\mathsf{t}_R^-(s+M-m+1),\nonumber\\
&&  {\rm for} \ s \pitchfork  x_L\le M-(m+1) \nonumber\\
&& \& \ 1\le m\le M-1, \nonumber\\
\mathscr{W}(s,s)&=&-(\mathsf{t}_L^+(s)+\mathsf{t}_L^-(s)\nonumber\\
&& +\mathsf{t}_R^+(s)+\mathsf{t}_R^-(s)),\nonumber\\
&& {\rm for} \ s \pitchfork m\ge 2.
\label{eq:W}
  \eea
We have introduced the notation $s\pitchfork $ with the meaning ``$s$
is to be taken for''.  The positive terms above correspond to jumps
{\em to} the state $\{x_L,m\}$, while the negative terms correspond to
jumps {\em from} the state $\{x_L,m\}$, see Figs.~\ref{fig:rates} and
\ref{fig:lattice}. The probability for a bubble of size $m=1$ is
altered by exchange with the $m=2$ state, or the $m=0$ ground state:
  \bea
\mathscr{W}(0,x_L+1)&=&\mathsf{t}_G^+(x_L), {\rm for} \ s
\pitchfork m=1,\nonumber\\
\mathscr{W}(s,s)&=&-(\mathsf{t}_G^-(x_L) +\mathsf{t}_L^-(s)+
\mathsf{t}_R^+(s)),\nonumber\\
& & {\rm for} \ s \pitchfork m=1.
\label{eq:W2}
  \eea
Finally, for the ground state population, we find
  \bea
\mathscr{W}(0,x_L+1)&=&t_G^-(x_L),\,\,\mbox{for } x_L\le M-1\nonumber\\
\mathscr{W}(0,0)&=&-\sum_{x_L=0}^{M-1} t_G^+(x_L),
\label{eq:W3}  
  \eea
i.e., the $m=0$ state can change by jumping to this state from $m=1$
(first term) or by jumping out of the $m=0$ state (second term).
There are $M$ possible jumps out from or to the ground state,
corresponding to bubble opening or closing at any of the $M$ internal
bps. The remaining matrix elements are equal to zero. The problem at
hand is that of determining the
eigenvalues and eigenvectors of the $(S+1)\times(S+1)$-matrix
$\mathscr{W}$ above. In terms of the running variable $s$, see
Eq.~(\ref{eq:s}), and the $\mathscr{W}$-matrix defined in
Eqs.~(\ref{eq:W}), (\ref{eq:W2}) and (\ref{eq:W3}) the detailed
balance conditions (\ref{eq:det_balance}) can be written as
  \be
\label{detail}
\mathscr{W}(s,s')P_{\rm eq} (s')=\mathscr{W}(s',s) P_{\rm eq} (s).
  \ee
The eigenvectors are orthonormal in the sense \cite{van_Kampen}
\begin{equation}
\label{ortho}
\sum_s\frac{Q_p(s)Q_{p'}(s)}{P^{\rm eq}(s)}=\delta_{p,p'}.
\end{equation}
Convenient checks of the numerical results then include: (i) there
should be one zero eigenvalue $\eta_0=0$, the corresponding eigenvector
is the equilibrium distribution, i.e. $Q_0(s)=P^{\rm eq}(s)$; (ii) the
remaining eigenvalues should be real and negative (so that $\eta_p>0$ for
$p\ge 1$); (iii) The eigenvectors should satisfy the orthonormality
relation, Eq.~(\ref{ortho}). Instead of working with the asymmetric matrix
$\mathscr{W}(s,s')$, for numerical purposes it is sometimes preferable to
use the symmetric matrix $\mathscr{V}(s,s')=\mathscr{Z}(s)^{-1/2}\mathscr{W}
(s,s')\mathscr{Z}(s')^{1/2}$, see Refs.~\cite{van_Kampen,gardiner} for details.
Indeed, the Matlab code we used to numerically solve the master equation,
is based on the $\mathscr{V}$-matrix.

\subsection{Survival and waiting time densities}

In this section we derive the expression for the waiting and survival time
densities given in Eqs.~(\ref{eq:Psi-Phi}).

\begin{figure}
\begin{center}
\includegraphics[width=6.8cm]{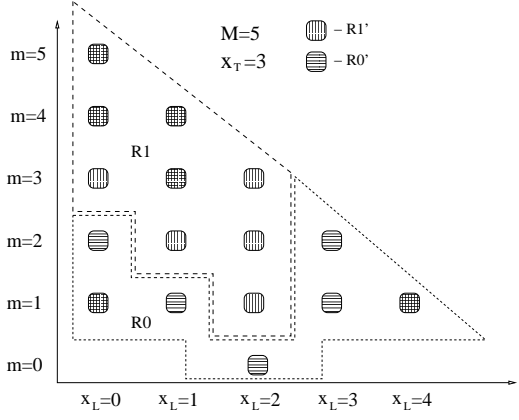}
\end{center}
\caption{Schematic of the $(x_L,m)$-points, region $\mathbb{R}$1
($\mathbb{R}$0), for which the stochastic variable takes the value
$I=1$ ($I=0$). The boundary points regions $\mathbb{R}1'$ and $\mathbb{R}0'$
are also indicated. The illustration is for the case $M=5$ and $x_T=3$, with
$\Delta=0$.}
\label{fig:lattice_R0_R1}
\end{figure}

Denote by $\rho(t|s_{\rm init})$ the first passage time density
starting from some initial position $s_{\rm init} \in \mathbb{R}1'$ or
$\mathbb{R}0'$, see
Fig. \ref{fig:lattice_R0_R1}. The survival time density $\phi(t)$ and
waiting time density $\psi(t)$ are then given by $\sum_{s_{\rm init}}
\rho(t|s_{\rm init})f (s_{\rm init})$, where $f (s_{\rm init})$ is the
distribution of initial points $\in \mathbb{R}1'$ (for $\phi (t)$) or
$\mathbb{R}0'$ (for
$\psi (t)$). Following standard arguments (see, e.g, \cite{translo}) we can
write the expression for $\rho(t|s_{\rm init})$, and therefore
$\psi(t)$ and $\phi(t)$, which becomes:
\begin{subequations}
  \bea
\label{eq:rho_ta}
\psi(t)&=&\sum_{p\in \mathbb{R}0} \bar{\eta}_p \bar{c}_p \exp(-\bar{\eta}_pt),\\
\phi(t)&=&\sum_{p\in \mathbb{R}1} \bar{\eta}_p \bar{c}_p \exp(-\bar{\eta}_pt),
\label{eq:rho_t}
  \eea
\end{subequations}
where 
  \be
\bar{c}_p=\sum_{s_{\rm init}} \frac{\bar{Q}_p (s_{\rm init}) f(s_{\rm init})}
{P_{\rm eq}(s_{\rm init})} \sum_{s} \bar{Q}_p (s).
\label{eq:c_p_surv_time}
  \ee
Here, $s\in \mathbb{R}0 (\mathbb{R}1)$ for $\psi(t)$ ($\phi(t)$), and
$\bar{\eta}_p$ and $\bar{Q}_p (s)$ are determined through the eigenvalue
equation (\ref{eq:eigenvalue_eq_reduced2}), which explicitly becomes in
$s$-space \cite{van_Kampen}
  \be
\sum_{\tilde{s}} \mathscr{W}(s,\tilde{s})\bar{Q}_p(\tilde{s})=
-\bar{\eta}_p \bar{Q}_p(s),
\label{eq:W_reduced}
  \ee
where $s,\tilde{s} \in \mathbb{R}1$ when calculating the survival time density, and
$s,\tilde{s} \in \mathbb{R}0$ for the waiting time density.

The problem is now reduced to obtaining the distribution of initial
points $f (s_{\rm init})$ such that agreement with the Gillespie time
series (see Appendix \ref{sec:Gillespie})
is obtained for long times. We define the rate coefficients for
jumps from the points in the boundary region $\mathbb{R}$1' to $\mathbb{R}$0'
(see Figs.~\ref{fig:lattice} and \ref{fig:lattice_R0_R1}): $t_{1\rightarrow 0}
(s_{\rm init})=t_L^+,$ $t_R^-,$ or $t_G^-$, where $s_{\rm init} \in
\mathbb{R}1'$.
  %\bea
%\mathsf{t}_{1\rightarrow 0}(s_{\rm init})&=&\mathsf{t}_L^+(s)\ {\rm for} \ s \pitchfork x_L=x_T-1, \nonumber\\ 
%& & \ m \in [2, M-x_T+1],\nonumber\\
%&=&\mathsf{t}_R^-(s)\ {\rm for} \ s \pitchfork x_L=x_T-m,\nonumber\\
%& & \ m \in [2, x_T],\nonumber\\
%&=&\mathsf{t}_G^-(s)\ {\rm for} \ s \pitchfork x_L=x_T-1,\nonumber\\ 
%& & \ m=1.
  %\eea
Similarly, for jumps from the points in the boundary region $\mathbb{R}$0' to
$\mathbb{R}1'$ ($s_{\rm init} \in \mathbb{R}0'$) we define:
$t_{0\rightarrow 1}(s_{\rm init})=t_L^-,$ $t_R^+,$ or $t_G^+$.
 %\bea
%\mathsf{t}_{0\rightarrow 1}(s_{\rm init})&=&\mathsf{t}_L^-(s)\ {\rm for} \ s \pitchfork x_L=x_T,\nonumber\\ 
%& &  \ m \in [1, M-x_T],\nonumber\\
%&=&\mathsf{t}_R^+(s)\ {\rm for} \ s \pitchfork x_L=x_T-1,\nonumber\\
%& & \ m \in [1 ,x_T-1],\nonumber\\
%&=&\mathsf{t}_G^+(s)\ {\rm for} \ s \pitchfork x_L=x_T-1,\nonumber\\ 
%& & \ m=0.
  %\eea
From the detailed balance condition we have that
  \be
\mathsf{t}_{1\rightarrow 0}(s) P_{\rm eq}(s)=\mathsf{t}_{0\rightarrow 1}(s')
P_{\rm eq}(s'),
\label{eq:det_balance_reduced}
  \ee
where $s$ and $s'$ are points in region $\mathbb{R}$1' and $\mathbb{R}$0'
which are
connected. For the the survival time density we then choose the
distribution of initial points proportional to the stationary influx
from region $\mathbb{R}0$'. Furthermore using the detailed balance
condition and normalizing we have for the initial distribution in the
$I=1$ state
  \be
f(s_{\rm init})= \frac{\mathsf{t}_{1\rightarrow 0}(s_{\rm init}) P_{\rm
eq}(s_{\rm init})}{\sum_{s_{\rm init}}\mathsf{t}_{1\rightarrow 0}(s_{\rm init}) P_{\rm
eq}(s_{\rm init})}\label{eq:f_sinit1}
  \ee
Similarly for the initial distribution in the I=0 state:
  \be
f(s_{\rm init})= \frac{\mathsf{t}_{0\rightarrow 1}(s_{\rm init}) P_{\rm
eq}(s_{\rm init})}{\sum_{s_{\rm init}}\mathsf{t}_{0\rightarrow 1}(s_{\rm init})
P_{\rm eq}(s_{\rm init})},
\label{eq:f_sinit2}
  \ee
which, together with Eqs.~(\ref{eq:rho_ta}), (\ref{eq:rho_t}) and
(\ref{eq:c_p_surv_time}),
determines $\psi(t)$ and $\phi(t)$. We below proceed to show the 
choices above for $f(s_{\rm init})$ satisfy ergodicity requirements.
 
Ergodicity requires that the ratio of times spent in the I=1 and I=0 state 
equals
  \be
R_{\rm eq}=\frac{\sum_{s\in \mathbb{R}1} P_{\rm eq} (s)}{\sum_{s\in
\mathbb{R}0} P_{\rm eq} (s)}.
  \ee
From Eq.~(\ref{eq:rho_t}) we have that the mean survival time can be written according to:
  \be
\tau_{\rm surv}=\int_0^\infty t \phi(t) dt = \sum_p (\bar{\eta}_p)^{-1}
\bar{c}_p,
\label{eq:tau_surv}
  \ee
and identically for the  mean waiting time $\tau_{\rm wait}$.
We proceed by noticing that the eigenvalue equation (\ref{eq:W_reduced})
can be written as
  \be
\sum_{\tilde{s}} \left(\mathscr{W}^{\rm refl}(s,\tilde{s})-\mathscr{W}^{\rm
abs}(s,\tilde{s})\right) \bar{Q}_p(\tilde{s})=-\bar{\eta}_p \bar{Q}_p(s),
  \ee
where 
  \be
\mathscr{W}^{\rm abs}(s,\tilde{s})=\mathsf{t}_{1\rightarrow 0}(s) \delta_{s,
\tilde{s}} \delta_{s,s_{\rm init}},
  \ee
with $s_{\rm init}\in \mathbb{R}1'$, and $\mathscr{W}^{\rm
refl}(s,\tilde{s})$ satisfy $\sum_s \mathscr{W}^{\rm
refl}(s,\tilde{s})=0$. Summing Eq.~(\ref{eq:W_reduced}) over $s$
and using the above identity we obtain
  \be
\sum_{s}\bar{Q}_p (s)=\sum_{s_{\rm init}} \bar{\eta}_p^{-1}
\mathsf{t}_{1\rightarrow 0}(s_{\rm init}) \bar{Q}_p (s_{\rm init})
\label{eq:relation_bulk_boundary}
  \ee
which is a useful connection between quantities in the bulk ($s \in
\mathbb{R}1$) and at the boundary $s_{\rm init}\in \mathbb{R}1'$. Applying
this relation to the expressions for the survival time,
Eqs.~(\ref{eq:c_p_surv_time}), (\ref{eq:f_sinit1}), and (\ref{eq:tau_surv}),
we find
  \be
\tau_{\rm surv}=\frac{\sum_p \sum_s \sum_{\tilde{s}}\bar{Q}_p (s)\bar{Q}_p
(\tilde{s})}
{\sum_{s_{\rm init}} \mathsf{t}_{1\rightarrow 0}(s_{\rm init}) P_{\rm eq}
(s_{\rm init})}.
  \ee
Finally, using the completeness relation \cite{van_Kampen} 
  \be
\sum_p\frac{\bar{Q}_p(s)\bar{Q}_p (\tilde{s})}{P_{\rm eq}(\tilde{s})}=
\delta_{s,\tilde{s}},
\label{eq:completeness}
  \ee
we see that 
  \be
\tau_{\rm surv}=\frac{\sum_{s\in \mathbb{R}1} P_{\rm eq} (s)}{\sum_{s_{\rm init}} \mathsf{t}_{1\rightarrow 0}(s) P_{\rm eq} (s_{\rm init})}.
  \ee
In a similar fashion
  \be
\tau_{\rm wait}=\frac{\sum_{s\in \mathbb{R}0} P_{\rm eq} (s)}{\sum_{s_{\rm
init}} \mathsf{t}_{0\rightarrow 1}(s) P_{\rm eq} (s_{\rm init})},
  \ee
which, using the detailed balance condition,
Eq.~(\ref{eq:det_balance_reduced}), shows that $\tau_{\rm
  surv}/\tau_{\rm wait}=R_{\rm eq}$.

With the completeness relation (\ref{eq:completeness}) and
Eq.~(\ref{eq:relation_bulk_boundary}) we find from
Eqs.~(\ref{eq:c_p_surv_time}), (\ref{eq:f_sinit1}), and (\ref{eq:f_sinit2})
that $\bar{c}_p$ can be written
  \be
\bar{c}_p=\frac{\bar{\eta}_p [\sum_{s} \bar{Q}_p (s)]^2}{\sum_p \bar{\eta}_p [\sum_{s} \bar{Q}_p (s)]^2}
  \ee
which is the form given in Eq.~(\ref{eq:bar_c_p_main}).

\begin{figure}
\begin{center}
\includegraphics[width=8.8cm]{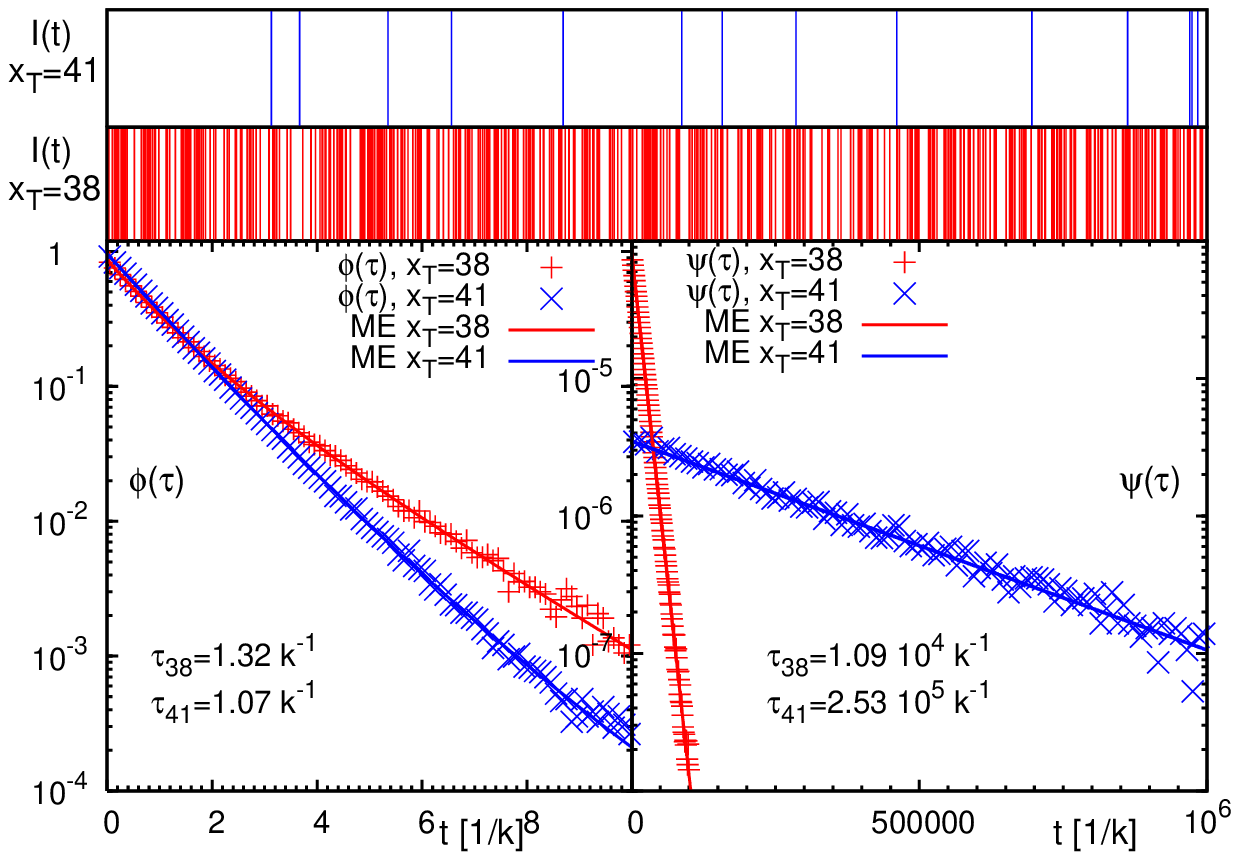}
\end{center}
\caption{Top: Fluorescence time series $I(t)$ for the T7 promoter
sequence, with tag position $x_T=38$ (red) and $x_T= 41$
(blue). Bottom: Waiting time ($\psi(\tau)$) and fluorescence survival time
($\phi(\tau)$) densities, in units of $k$. The data points (solid lines) are
results from the Gillespie algorithm (master equation).
All results are for $T=37^\circ C$ and 100 mM NaCl with DNA parameters
from \cite{FK}. }
\label{signal}
\end{figure}

We show in Fig.~\ref{signal} the time series obtained from a stochastic
simulation (see Appendix \ref{sec:Gillespie} for a short introduction, and
refer to Ref.~\cite{suman} for details) for two different tag positions in 
the T7 bacteriovirus promoter sequence
\begin{equation}
\begin{array}{l}
\mbox{\small\texttt{\textcolor{white}{AAAA}1\textcolor{white}{%
AAAAAAAAAAAAAAAAAA}20}}\\[-0.15cm]
\mbox{\small\texttt{\textcolor{white}{AAAA}|\textcolor{white}{%
AAAAAAAAAAAAAAAAAA}|\textcolor{white}{AAAAAAAAA}\textcolor{white}{AAA}}}\\[-0.15cm]
\mbox{\small\texttt{5'-aTGACCAGTTGAAGGACTGGAAGTAATACGACTC}}\\
\mbox{\small\texttt{\textcolor{white}{AAA}AG}\textcolor{red}{
\texttt{TATA}}\texttt{GGGACAATGCTTAAGGTCGCTCTCTAGGAg-3'}}\\[-0.15cm]
\mbox{\small\texttt{\textcolor{white}{AAAAAAA}|\textcolor{white}{AA}|
\textcolor{white}{AAAAAAAAAAAAAAAAAAAAAAAAA}|\textcolor{white}{AAA}}}\\[-0.15cm]
\mbox{\small\texttt{\textcolor{white}{AAAAAAA}\textcolor{red}{38}%
\textcolor{white}{A}\textcolor{blue}{41}\textcolor{white}{%
AAAAAAAAAAAAAAAAAAAAAAAAA}68\textcolor{white}{AAA}}}
\end{array}
\end{equation}
whose TATA motif is marked in red \cite{Kalosakas_Rasmussen}. A promoter is
a sequence (often containing the so called TATA motif) marking the start of
a gene, to which RNA polymerase is recruited and where transcription then
initiates. Fig.~\ref{signal} shows the signal $I(t)$ at $37^{\circ}$C for the
tag positions $x_T=38$ in the core of TATA, and $x_T=41$ at the second GC bp
after TATA. Bubble events occur
much more frequently in TATA (the TA/AT stacking interaction is
particularly weak \cite{FK}). This is quantified by the density of
waiting times $\psi(\tau)$ spent in the $I=0$ state, whose
characteristic time scale $\tau'=\int_0^\infty d\tau \tau \psi(\tau)$
is more than an order of magnitude longer than at $x_T=41$. In contrast, we
observe similar behavior for the density of opening times $\phi(\tau)$ for
$x_T=38$ and $41$. The solid lines are the results from the master equation,
see subsection \ref{sec:ME_surv_wait}, showing excellent agreement with
the Gillespie results. Notice that whereas $\psi(t)$ is characterized
by a single exponential, $\phi(t)$ show a crossover between different
regimes. For long times both $\psi(\tau)$ and $\phi(\tau)$ decay
exponentially as it should for a finite DNA stretch.

\section{Reduced one-variable scheme for a homopolymer}
\label{sec:one_variable}

After addressing the derivation of the probability densities $\psi(t)$ and
$\phi(t)$, and the details concerning the transfer matrix $\mathscr{W}$, we
show how the master equation formalism reduces when a homopolymer sequence
is considered, that is, a sequence with only one type of bps such as $(AT)_N$.
Homopolymers can be realized experimentally. In the case they are clamped,
possible secondary structure formation does not appear to occur, and our
formalism remains valid. In the case of long homopolymers, imperfect matching
conditions apply, and additional degrees of freedom emerge \cite{poland}.
Although this can be straightforwardly included in the formalism, we do
not consider this case here.

In the homopolymer case, it is possible to obtain analytical results. To
that end we note that for a homopolymer, all bps have the same statistical
weights $u_{\mathrm{st}}(x)$ and $u_{\mathrm{hb}}(x)$. Formally, we
therefore use $u=u_{\mathrm{st}}u_{\mathrm{hb}}$ for disruption of
additional bps after bubble initiation. Due to this choice, we need to
utilize the initiation factor $\sigma_0$ instead of the ring factor $\xi$,
as $\sigma_0$ takes care of the fact that upon initiation two stacking
bonds are broken \cite{blake,santalucia,FK}.
If we furthermore assume that we are below the melting
temperature $u<1$, have a long DNA region $M\gg 1$ and consider
bubbles far from the clamping, end effects are much less pronounced.  It
then follows that $P(x_L,m,t;x_L',m')=\tilde{P}(m,t; m')$, and the
master equation reduces to
  \bea
\frac{\partial}{\partial t}\tilde{P}(m,t)&=&\tilde{\mathsf{t}}^+(m-1)
\tilde{P}(m-1,t)\nonumber\\
\nonumber
&&+\tilde{\mathsf{t}}^-(m+1)\tilde{P}(m+1,t)\\
&&-(\tilde{\mathsf{t}}^+(m)+\tilde{\mathsf{t}}^-(m) ) \tilde{P}(m,t),
\label{eq:master_eq_reduced}
  \eea
with the short-hand notation $\tilde{P}(m,t)=\tilde{P}(m,t; m')$.
The forward transfer coefficients in this limit are given by 
 \bea
\tilde{\mathsf{t}}^+(m=0)&=&k \sigma_0 u s(0)\nonumber\\
\tilde{\mathsf{t}}^+(m)|_{m\ge 1}&=&kus(m),
  \eea
where we have incorporated the fact that a bubble size increase
can occur by opening of a bp at either the left or the
right fork. For the backward transfer coefficients, we find
\be
\tilde{\mathsf{t}}^-(m)=k.
 \ee
The eigenvalue equation corresponding to Eq.~(\ref{eq:master_eq_reduced})
has the comparatively simple structure
 \bea
\nonumber
&&\tilde{\mathsf{t}}^+(m-1)\tilde{Q}_p(m-1)\\
&&+\tilde{\mathsf{t}}^-(m+1)\tilde{Q}_p(m+1)\nonumber\\
&&-(\tilde{\mathsf{t}}^+(m)+\tilde{\mathsf{t}}^-(m) ) \tilde{Q}_p(m)=
-\tilde{\eta}_p \tilde{Q}_p(m),
\label{eq:eigenvalue_eq_reduced}
  \eea
with eigenvalues $\tilde{\eta}_p$ and eigenvectors $\tilde{Q}_p(m)$
($p=0,1,\ldots,M$). The equation above is identical to the one in 
\cite{PRL,JPC}, and thus our generalized formalism is consistent with
previous homopolymer models \cite{PRL,JPC}. We note that the
equilibrium distribution becomes
\be
\tilde{P}_{\rm eq}(m)=\frac{\mathscr{Z}(m)}{\sum_{m=0}^M \mathscr{Z}(m)},
\ee
where $\mathscr{Z}(m)=\sigma_0 (1+m)^{-c}u^m$
with $\mathscr{Z}(0)=1$, see Eqs.~(\ref{part}) and (\ref{eq:P_eq}).  

The autocorrelation function is, as before, 
simply proportional to the joint probability of having $m\ge 1$ at
time $t$ and $m\ge 1$ at initial time $t=0$. Proceeding as previously,
and assuming that initially the system is at equilibrium,
$P(m,0;m',0)=\delta_{mm'}\tilde{P}_{\rm eq}(m)$, we have
 \be
\tilde{A}(t)=\langle I(t)I(0)\rangle -(\langle I \rangle)^2= \sum_{p\neq 0}
\left(\tilde{T}_p\right)^2\exp\left(-\frac{t}{\tilde{\tau}_p}\right),
\label{eq:A_reduced}
  \ee
where $\tilde{T}_p=\sum_{m=1}^{M}\tilde{Q}_p(m)$. Here, we introduced
the relaxation times $\tilde{\tau}_p\equiv 1/\tilde{\eta}_p$.  
As before, we write the correlation function according to
$\tilde{A}(t)=\int d\tau \exp(-t/\tau) \tilde{f}(\tau)$, with the
relaxation time spectrum
\be
\tilde{f}(\tau)=\sum_{p\neq 0} \left( \tilde{T}_p\right)^2 \delta(\tau-
\tilde{\tau}_p).
\label{eq:f_tau_reduced}
\ee

\begin{figure*}
\begin{center}
\includegraphics[width=5.8cm]{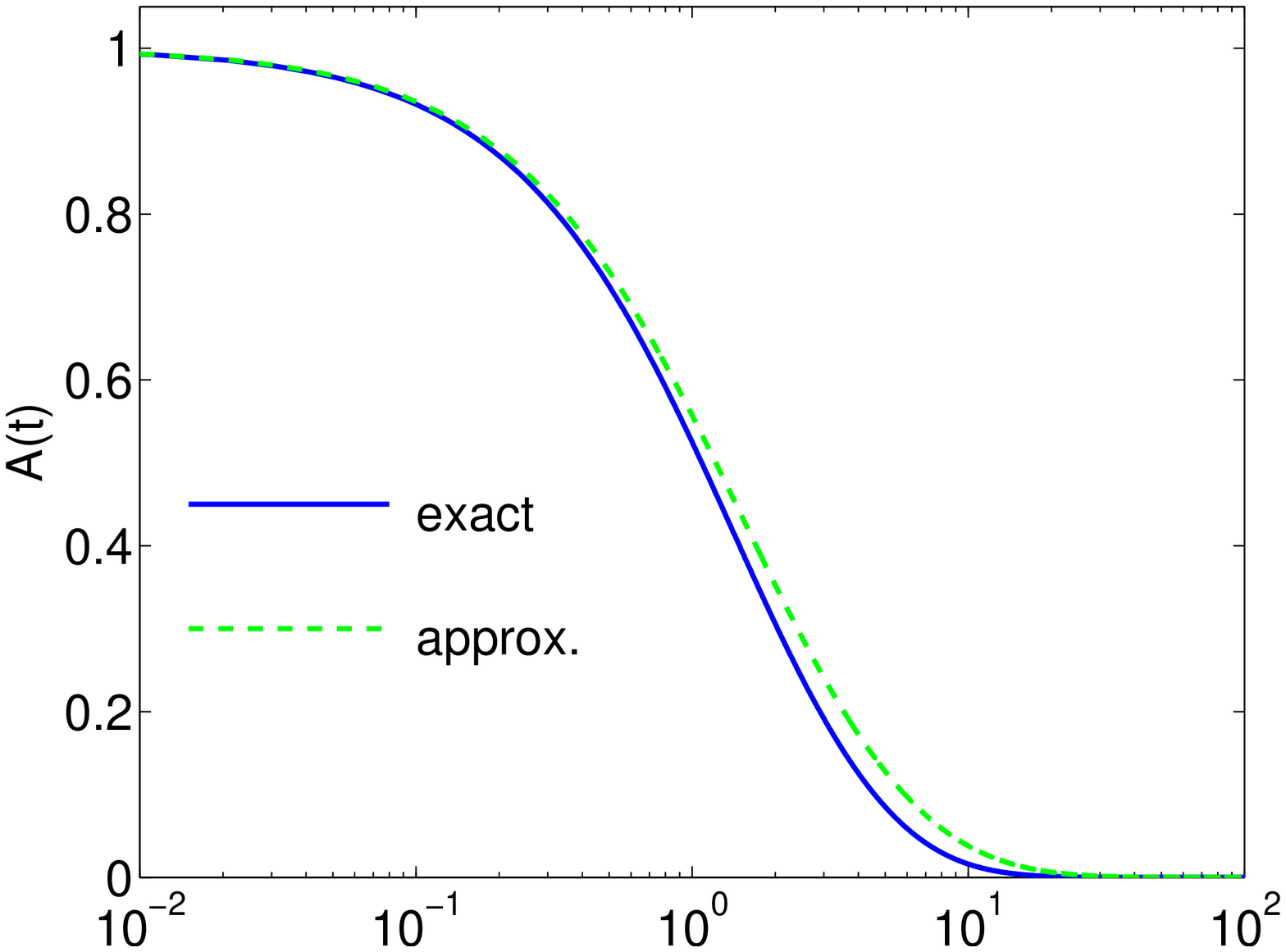}
\includegraphics[width=5.8cm]{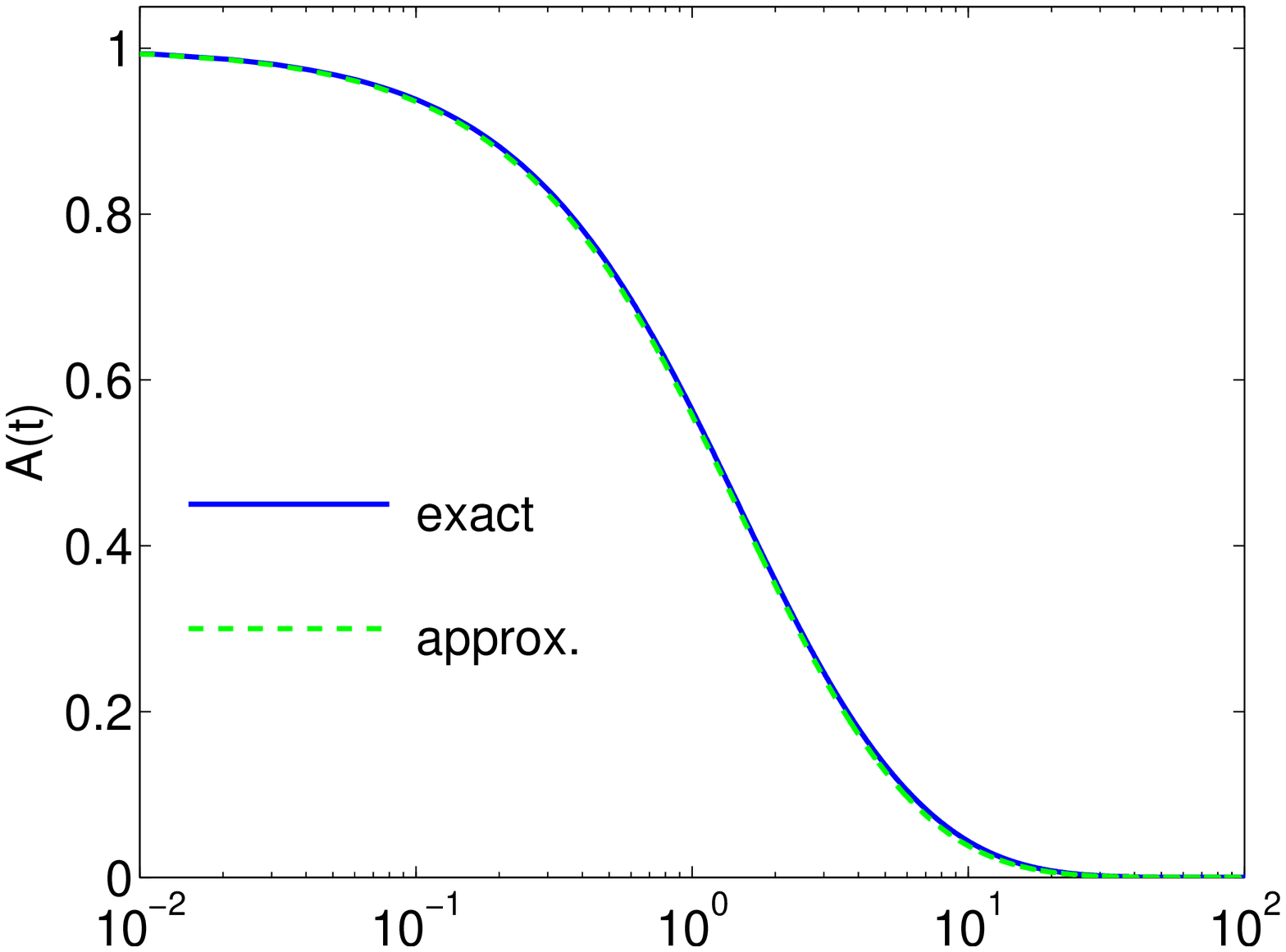}
\includegraphics[width=5.8cm]{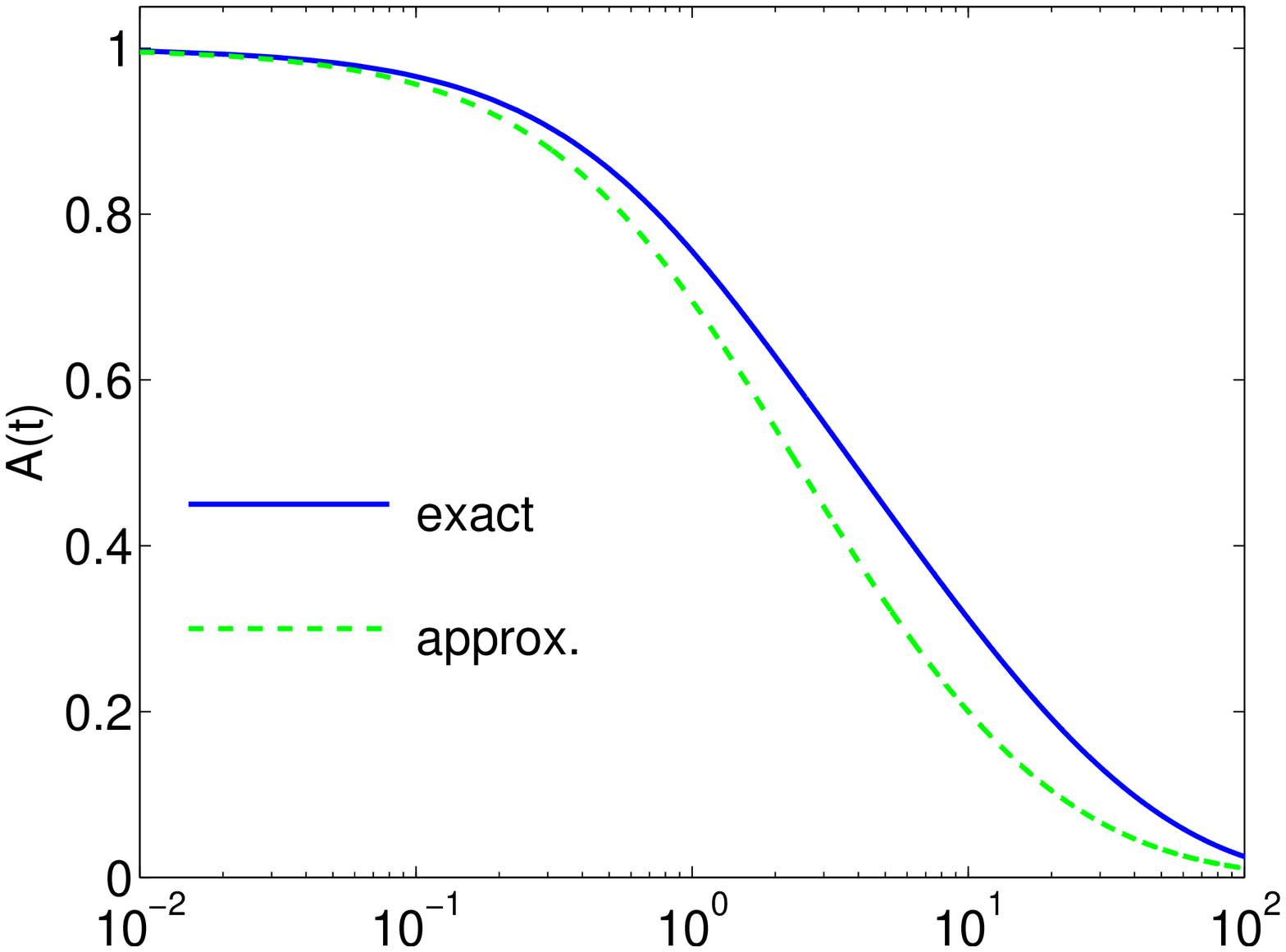}\\
\includegraphics[width=5.8cm]{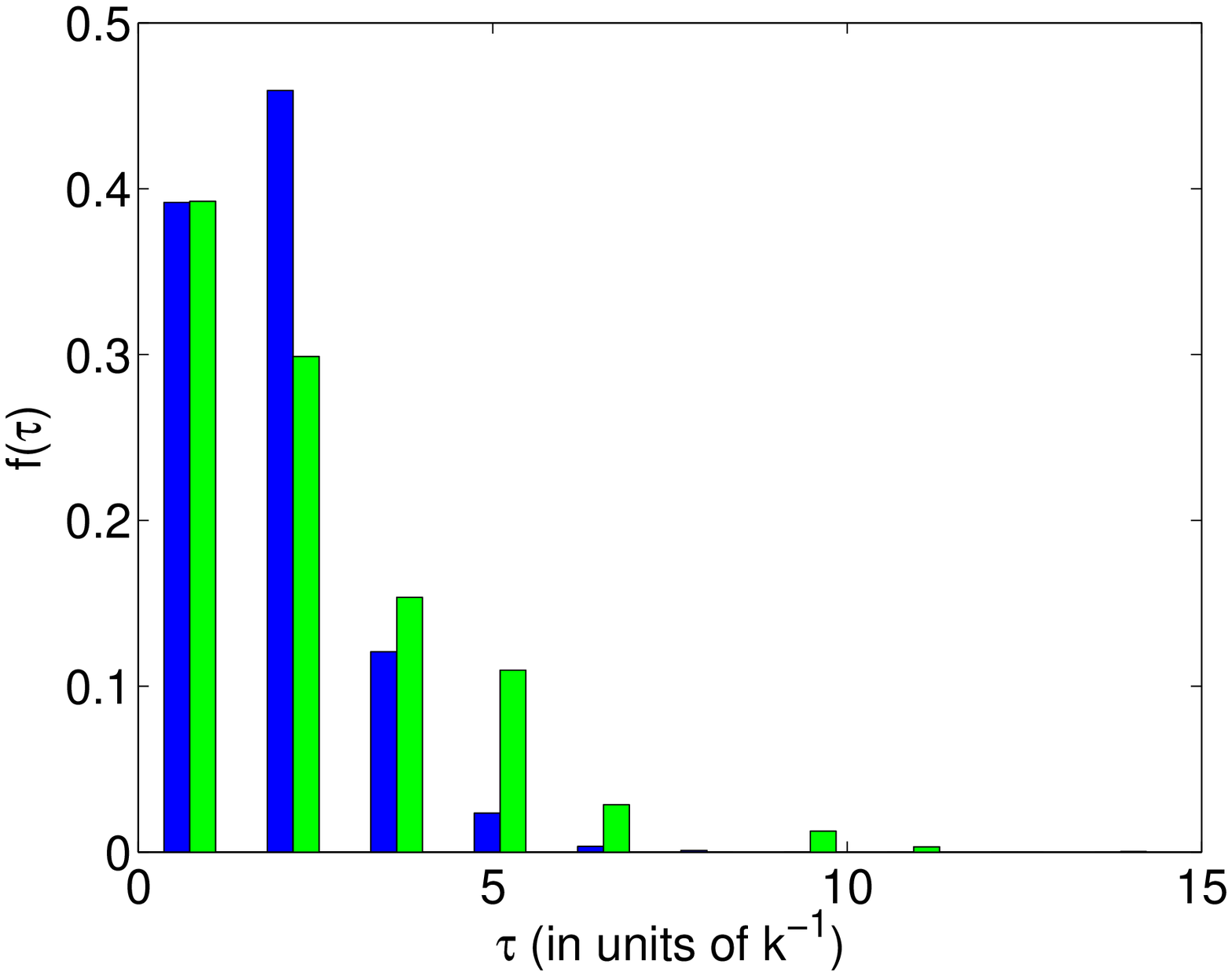}
\includegraphics[width=5.8cm]{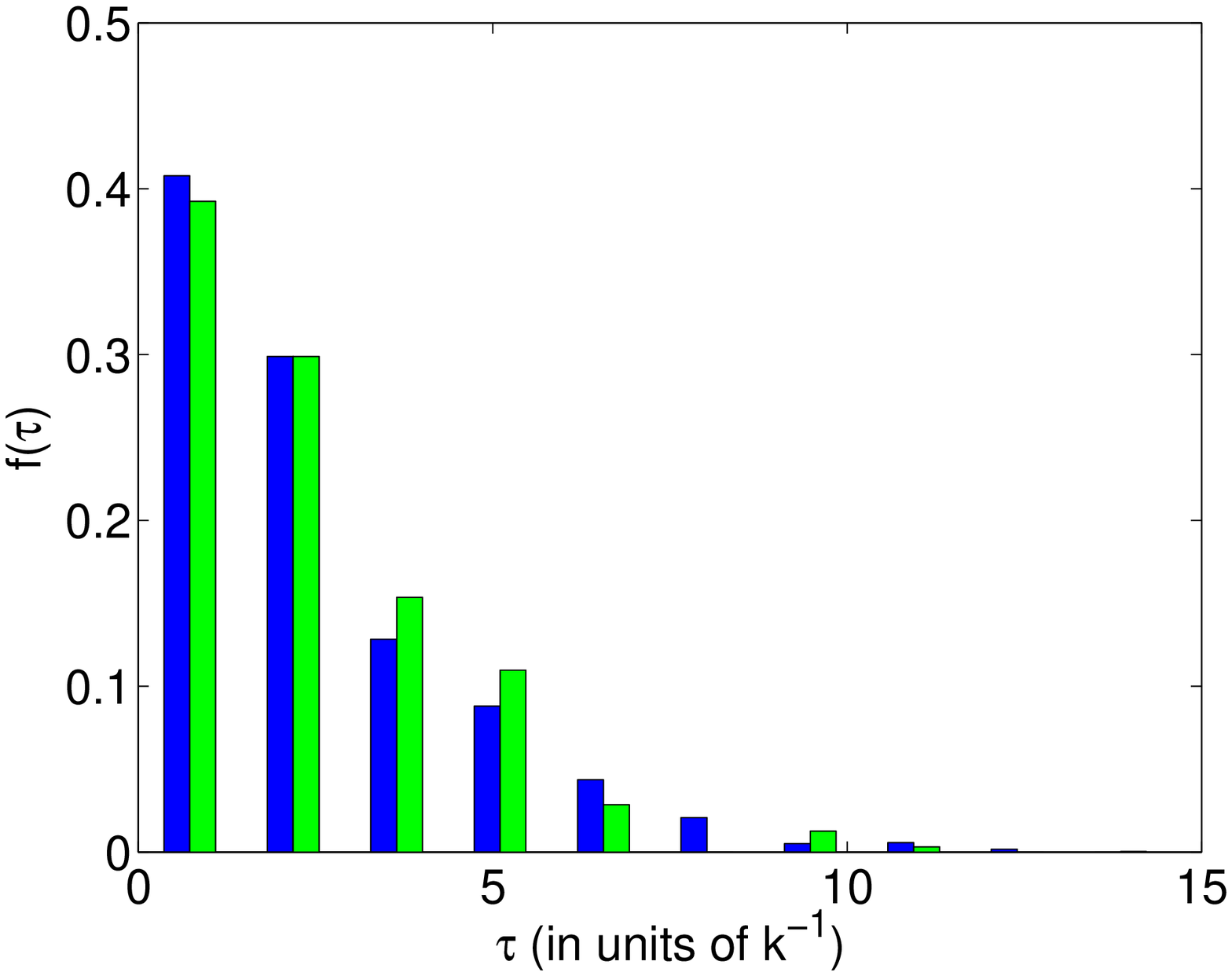}
\includegraphics[width=5.8cm]{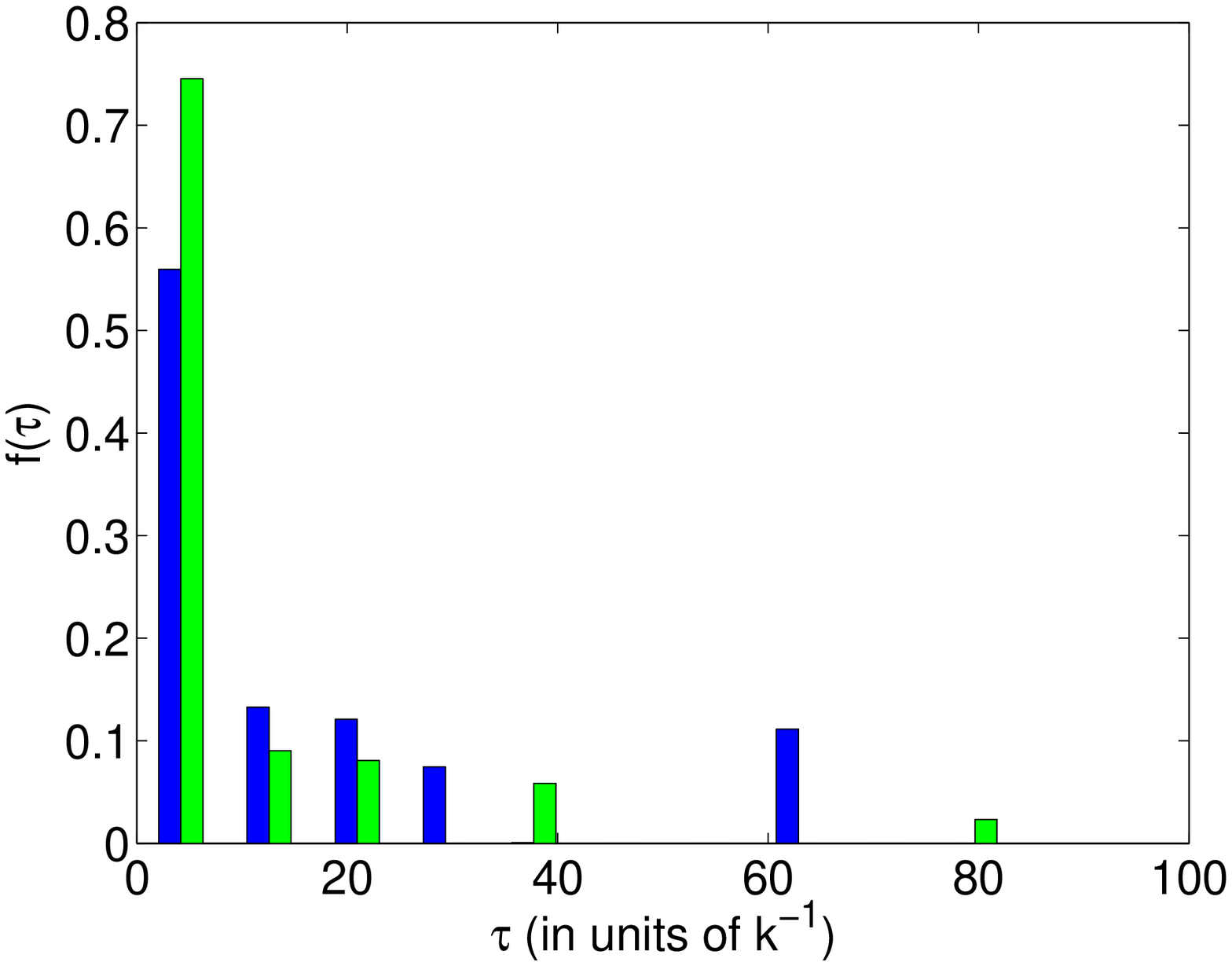}
\end{center}
\caption{Autocorrelation $A(t)$ and spectral density $f(\tau)$ for a
  tagged bp in a homopolymer region: $u=u_{\rm hb}$$u_{\rm st}$.
  Left: Close-to-end-tagging far from $T_m$ ($x_T=2$,
  $u=0.6$). Middle: Center-tagging far from $T_m$ ($x_T=20$,
  $u=0.6$). Right: Center-tagging close to $T_m$ ($x_T=20$,
  $u=0.9$). In the $A(t)$ plots the blue curves are the exact
  result. The dashed green curves are approximated from from
  Eqs.~(\ref{eq:eigenvalue_eq_reduced}) and (\ref{eq:A_reduced}). In the
  spectral density plot the data were collected into 10 bins. The
  green bars are the approximate one-variable results,
  Eq.~(\ref{eq:f_tau_reduced}), and the blue bars are the exact
  result. The length of the DNA segment was $M=40$.
  The approximate expression only works well for
  internal tagging and below $T_m$.}
\label{fig:A_t_homo}
\end{figure*}

In Fig.~\ref{fig:A_t_homo}, we compare the approximate result for
$A(x_T,t)$ obtained by numerical solution of
Eqs.~(\ref{eq:eigenvalue_eq_reduced}), and using
Eq.~(\ref{eq:A_reduced}), with the general result from the master equation in
Section \ref{sec:ME}. We also show the corresponding weighted spectral
densities given by Eq.~(\ref{eq:f_tau_reduced}). We note that the
approximate expression works well only for the case of internal
tagging and temperatures below the melting temperature (and for a
sufficiently long DNA region); for a short DNA sequence,
close-to-end-tagging or high temperatures (i.e., large bubbles) end
effects, which are not included in the approximate model above, are
significant.

\begin{figure}
\begin{center}
\includegraphics[width=8.8cm]{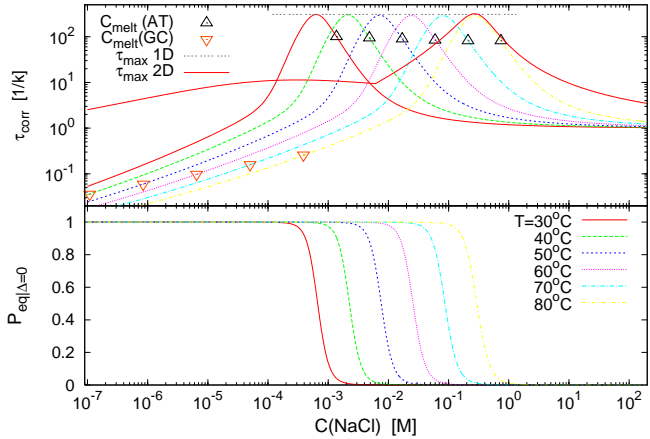}
\end{center}
\caption{Top: Mean correlation time $\tau_{\rm corr}=\int_0^\infty dt' A(t')/
A(0)$
versus NaCl concentration for various temperatures $T$ for the AT9 construct
of Ref.~\cite{altan}, showing a critical slowing down at the melting
concentration (compare lower panel). The
triangles denotes the melting concentration for infinitely long random
AT and GC stretches. The curve denoted by $\tau_{{\rm max} 1D}$ is the
result given in Eq.~(\ref{1drelax}), and $\tau_{{\rm max} 2D}=\max\{\tau_p\}$,
$p=1,\ldots,S$ is the maximum relaxation time of the full problem specified
in Sec. III.
Bottom: Opening probability of bp $x_T$.}
\label{salttemp}
\end{figure}

In the analysis of Refs.~\cite{PRL,bj} it was found that close
to the melting transition at $T_m$, the mean correlation
function takes its maximum (critical slowing down). In order to get an
understanding of this behavior we here analytically obtain the
largest relaxation time from the homopolymer model above. From
Reference \cite{tobias} we have that the eigenvalues, see
Eq.~(\ref{eq:eigenvalue_eq_reduced}), are for $c=0$
  \be
\tilde{\eta}_p =k (u+1-2 u^{1/2} \cos \omega_p)\label{eq:homo_eigenvalues}
  \ee
where $\omega_p$ ($0<\omega_p \le \pi$) is obtained from the
transcendental equation 
  \be
g(\omega_p)=\sin [(M+1)\omega_p]-\delta\sin [M\omega_p]=0
  \ee
with $\delta=(1-\sigma_0)u^{1/2}$. 
For $u\rightarrow 1$ and $\sigma_0\rightarrow 0$ we get
  \bea
g(\omega_p)&=&\sin [(M+1)\omega_p]- \sin [M\omega_p]\nonumber\\
&=&2 \sin \frac{\omega_p}{2} \cos [(M+\frac{1}{2})\omega_p]
  \eea
so that we have 
  \be
\omega_p=\frac{(p-1/2)\pi}{M+1/2}
  \ee
which together with Eq.~(\ref{eq:homo_eigenvalues}) give the
eigenvalues. The smallest eigenvalue (largest relaxation time) is
obtained for $p=1$, i.e.  $\tilde{\eta}_1=2k (1-\cos(\pi/[2M+1])) \approx k
\pi^2/(2M+1)^2$ for $M\gg 1$, and therefore the largest
relaxation time becomes
  \be
\label{1drelax}
\tilde{\tau}_1=\frac{1}{\tilde{\eta}_1}\approx \frac{(2M+1)^2}{\pi^2} k^{-1}
  \ee
We notice that the longest relaxation time scales as $\sim M^2$ at
melting, in agreement with the findings in \cite{bicout}. Fig.~\ref{salttemp}
demonstrates the good agreement of the homopolymer result ($\tau_{\mathrm{max
}}$, 1D in the figure) with the maxima of the correlation time, that coincide
with the melting concentration.

\section{Conclusions}

In this study we considered the bubble breathing dynamics in a
heteropolymer DNA-region characterized by statistical weights $u_{\rm
st}(x)$ for disrupting a stacking interaction between neighboring
bps, and the weight $u_{\rm hb}(x)$ for breaking a Watson-Crick hydrogen bond
($x$ labels different bps), as well the bubble initiation parameter (the
ring-factor) $\sigma_0$ ($\xi$). For that purpose, we introduced a
$(2+1)$-variable master equation governing the time evolution of the
probability distribution to find a bubble of size $m$ with left fork
position $x_L$ at time $t$, as well as a complementary Gillespie
scheme. The time averages from the stochastic simulation agree well
with the ensemble properties derived from the ME. We calculate the
spectrum of relaxation times, and in particular the experimentally
measurable autocorrelation function of a tagged bp is obtained.  All
parameters in our model are known from recent equilibrium measurements
available for a wide range of temperatures and NaCl concentrations, except for
the rate constant $k$ for (un)zipping that is the only free fit
parameter. A better understanding of the zipping rate $k$ remains an open
question, requiring a detailed microscopic modelling of DNA-breathing.

For the case of a long homopolymer DNA region with internal tagging
and below the melting temperature the position of the bubble becomes
negligible, and the master equation reduces to previous (1+1)-variable
approaches in terms of the bubble size.

\begin{acknowledgments}

We thank Maxim Frank-Kamenetskii, Oleg Krichevsky, and Kim Splitorff for
very helpful discussion.
SKB acknowledges support from Virginia Tech through ASPIRES
award program. TA acknowledges partial support through a Research Career
Award from the Knut and Alice Wallenberg foundation.
RM acknowledges partial funding from the Natural Sciences and Engineering
Research Council (NSERC) of Canada and the Canada Research Chairs program.

\end{acknowledgments}

\begin{appendix}

\section{Gillespie approach}
\label{sec:Gillespie}

In this section we briefly review the Gillespie algorithm. Together with the
explicit expressions for the transfer coefficients introduced in the
previous section it is used to generate stochastic time series of bubble
breathing. In particular we show how the motion of a tagged bp is
obtained.

To denote a bubble state of $m$ broken bps at position $x_L$ we
define the occupation number $b(x_L,m)$ for each lattice point in
Fig.~\ref{fig:lattice} with the properties $b(x_L,m)=1$ if the particular
state $\{x_L,m\}$ is occupied and $b(x_L,m)=0$ for unoccupied states. For
the completely zipped state $m=0$ there is no dependence on $x_L$, and we
introduce the occupation number $b(0)$. The stochastic DNA
breathing then corresponds to the nearest neighbor jump
processes in the triangular lattice in Fig. \ref{fig:lattice}.  Each
jump away from the state $\{x_L,m\}$ (i.e., from the state with
$b(x_L,m)=1$) occurs at a random time $\tau$ and in a random
direction to one of the nearest neighbors; it is governed by the
reaction probability density function \cite{Gillespie,Gillespie1}
\begin{equation}
\label{gill}
P(\tau,\mu,\nu)=\mathsf{t}^{\mu}_{\nu}(x_L,m)\exp\left(-\tau\sum_{\mu,\nu} \mathsf{t}^{\mu}_{\nu}(x_L,m)\right),
\end{equation}
which for a given state $(x_L,m)$ defines after what waiting time $\tau$ the
next step occurs and in what 'direction', $\nu\in\{G,L,R\}$, $\mu\in
\{+/-\}$. A simulation run produces a time series of occupied
states $\{x_L,m\}$ and how long time $\tau=\tau_j$ ($j=1,...,N$, where
$N$ is the number of steps in the simulation) this particular state is
occupied. This waiting time $\tau$, that is, according to Eq.~(\ref{gill})
follows a Poisson distribution \cite{cox}.

\subsection{Tagged bp survival and waiting time densities}

The stochastic
variable $I(t)$ is then obtained by summing the Gillespie occupation
number $b(x_L,m)$ ($b(x_L,m)$ takes only values $0$ or $1$) over
region $\mathbb{R}1$, i.e.
  \be
I(t) =  \sum_{x_L,m\in \mathbb{R}1} b(x_L,m).
  \ee
From the time series for $I(t)$ one can, for instance, calculate the
waiting time distribution $\psi(\tau)$ of times spent in the $I=0$
state, as well as the survival time distribution $\phi(\tau)$ of times
in the $I=1$ state. Explicit examples for $\psi(\tau)$ and
$\phi(\tau)$ are shown in Sec.~\ref{sec:ME_details}.

The probability that the tagged bp is
open becomes 
  \be
P_G(t_j)=\frac{1}{t_N}\sum_{j=1}^{N} \tau_j I(t_j)
  \ee
where $t_j=\sum_{j'=1}^{j} \tau_{j'} $. For long times the explicit
construction of the Gillespie scheme together with the detailed
balance conditions guarantee that $P_G(t_j)$ tends to the equilibrium
probability, i.e. that $P_G(t_j\rightarrow\infty)=\sum_{x_L,m\in \mathbb{R}1}
P^{\rm eq}(x_L,m)$, where $P^{\rm eq}(x_L,m)$ is given in
Eq.~(\ref{eq:P_eq}).

\subsection{Tagged base-pair autocorrelation function}

The autocorrelation function for a tagged bp is obtained through 
\begin{eqnarray}
\nonumber
A_t(x_T,t)&=&\overline{I(t)I(0)}-(\overline{I(t)})^2\\
&&\hspace*{-1.8cm}
=\frac{1}{T}\int_0^TI(t+t')I(t')dt'-\left(\frac{1}{T}\int_0^TI(t')dt'\right)^2
\label{eq:A_t_xT}
\end{eqnarray}
which for long times agrees with the the ensemble average, Eq.~(\ref{eq:A_t}),
from the master equation.  The function $A_t(x_T,t)$ corresponds to the 
blinking autocorrelation function obtained in the FCS experiment from
Ref.~\cite{altan}.

\end{appendix}


\begin{thebibliography}{99}
\bibitem{REM} Note that for $T=37^{\circ}$C we used that $1k_BT=0.62$kcal/mol.
Also note that writing Boltzmann factors for the free energies as $\exp(\beta
\Delta G)$, with $\beta=1/(k_BT)$, a positive $\Delta G$ denotes an unstable
bond.
\end{thebibliography}
\end{document}